\documentclass[11pt]{article} 

\title{Topological classifier for detecting the emergence \\of epileptic seizures}
\author{Piangerelli Marco, Matteo Rucco and Emanuela Merelli
\footnote{Computer Science Division, School of Science and Technology, University of Camerino, Camerino, Italy}}
\pdfoutput=1
\usepackage{amssymb}
\usepackage{pgf}
\usepackage{tikz}
\usepackage{amsmath}
\usepackage{framed}
\usepackage{rotating}

\usepackage{geometry}

\usetikzlibrary{arrows,automata}
\usetikzlibrary{positioning}

\begin{document}
\maketitle

\section*{Abstract}
\noindent In this work we study how to apply topological data analysis to create a method suitable to classify EEGs of patients affected by epilepsy. The topological space constructed from the collection of EEGs signals is analyzed by Persistent Entropy acting as a global topological feature for discriminating between healthy and epileptic signals. The Physionet data-set has been used for testing the classifier.

\bigskip
\noindent {\small Keywords:} {\scriptsize Complex systems, Brain, Epilepsy, Topological Data Analysis, Persistent Entropy, Classifier, Time series}

\medskip
\section{Introduction}

According to the World Health Organization (WHO), epilepsy is a chronic brain disorder characterized by recurrent seizures, which may vary from a brief lapse of attention or muscle jerks to severe and prolonged convulsions. The seizures are caused by sudden, usually brief, excessive electrical discharges in a group of brain cells (neurons)\footnote{www.who.int/topics/epilepsy/}. 
Epilepsy can be classified according to the portion of the involved neural cortex: the {\it{focal}} epilepsy seizures are spatially confined either in small part of a lobe or in the whole hemisphere, while the {\it{generalized}} epilepsy seizures spread in the whole brain. 
In both cases epileptic seizures are a spontaneous hyper-synchronous activity of clusters of neurons ~\cite{Majumdar2014}. It is worth mentioning that the human brain is a complex self adaptive system,  
composed of billion of non-identical neurons, entangled in loops of non-linear interactions, those determining the brain behaviors; the epilepsy is an example ~\cite{Telesford2011}.
In our perspective, identifying the onset of a neural hyper-synchronization is likely to discover patterns of information expressed by a network of interactions in the space of neurons. Unfortunately, to the best of our knowledge, monitoring the electrical activity of each single neuron is not feasible, 
but it is possible to capture the electrical activity of the whole brain (or of a part of it). Currently, the standard technique used to record the signals through the positioning of certain number of electrodes on the scalp, is the electroencephalogram (EEG), whose direct observation helps neurologists in diagnosing epilepsy; the use of methods for the automatic diagnosis is still far from be a reality. In the last decades several methods have been proposed in literature -- linear and non-linear analysis~\cite{Yang2008, McSharry2003}, applications of chaos theory~\cite{Iasemidis1996} and dynamical systems modelling~\cite{Santaniello2011,Iasemidis2001}, but none of them is suitable 
The intrinsic non-linearity and the non-stationarity of EEG signals request for more suitable methods capable of extracting global information, both structural and functional, characterizing the brain cluster of neurons involved in the hyper-synchronous activity. 

To this end, starting from the analysis of the EEG signals we must reconstruct the abstract structural component that represents the interactions among neuronal clusters involved in the brain activity and from the analysis of the evolution of the structural component (the abstract model of the brain) based on windows of observations, we must be able to identify the emergence of an hyper-synchronous activity of the brain neurons. In a short communication~\cite{epilepsy}, the authors have proposed the use of topological data analysis (TDA for short) for EEG signals, to characterize the structural component and the emergence of epileptic seizures.  This approach based on TDA  has been previously introduced within the TOPDRIM\footnote{Topology driven methods for complex systems: www.topdrim.eu} project for the analysis of fMRI signals~\cite{petri2014homological}.  In this paper we describe a method suitable to automatically classify EEG signals that record epileptic seizures from those recording the activity of an healthy brain. This step is very important and preliminary to reconstruct the abstract structural and functional components of the brain because it allows us to characterize the transition phase from an healthy to an epileptic state of the system.
The proposed method uses the topological data analysis (TDA) combining two powerful instruments, persistent homology and persistent entropy measure for analysing the set of EEG signals and construct a topological classifier. 
TDA consists of a set of algorithms for investigating the higher dimensional relations hidden in data sets through the construction of simplicial complex; it finds its theoretical justification in the branch of mathematics called topology. The key-concept in TDA is the concept of \textit{persistent homology}: a procedure for counting, through a process called \textit{filtration}, the higher dimensional persistent holes. TDA has been applied in various fields among which computer vision and sensor coverage problem ~\cite{de2007coverage}, signal analysis~\cite{perea2015sliding,tdaLoccioni}, biological problems ~\cite{chan2013topology,ibekwe2014topological,petri2014homological} and in studying the spreading of contagions in social and biological systems~\cite{taylor2015topological}. Persistent Entropy is a Shannon-like entropy for the measurement of the information discovered during the filtration process within the TDA. Shannon entropy ~\cite{rosso2001wavelet} is the well known concept in thermodynamics as well as in information theory. The contribution of this work is twofold: from the one hand, we describe how the Persistent Entropy (PE for short) can be used to discriminate the epileptic state versus non epileptic states. From the other hand, we conjecture that the Vietoris-Rips filtration helps to understand which region plays the role of trigger for an epileptic seizure.
The paper is organized as it follows: in Section ~\ref{sec:new_paradigm} we introduce the new paradigm of TDA and in Section ~\ref{sec:classifier} the TDA-based methodology used for the implementation of the new classifier is explained; finally in Section ~\ref{sec:results} the results of the work are presented and discussed.

\section{TDA: a new paradigm for data analysis}
\label{sec:new_paradigm}
In this section we introduce the TDA paradigm as the tool for analysing the EEG data set and discovering elements that are correlated in a given a data-set by $n$-relations.  Suppose we have a set of points $G$, i.e. data, embedded in a $d$-dimensional space $\mathbb{D}^d$ and assume that those data were sampled by an unknown $k$-dimensional space $\mathbb{V}^k$ with $k\leq d$, our task is to reconstruct the space $\mathbb{V}^k$ from the data set $G$, as it is in solving the inverse problem. 

For instance, traditional analysis techniques, such as manifold learning and principal components analysis (PCA)~\cite{nasuti2016metal}, or unsupervised clustering techniques~\cite{domingues2016pyrethroid}, make some fundamental assumptions over the space $\mathbb{V}^k$: it is a manifold; it is locally Euclidean; its metric is well defined; it lacks of curvature but it is smooth ~\cite{zomorodian2007topological}.
Since real data-sets are not a vector space they violate those assumptions, as consequence we claim that $\mathbb{V}$ and $\mathbb{D}$ have to be reshaped to \textit{topological spaces}. Intuitively a topological space is a set of elements that are equipped with a notion of \textit{proximity} parameter that gives rise to a coordinate-free metric. Instead of using the points as the vertices of a combinatorial graph to capture local information, in TDA the data set is converted into a topological space to discover the higher dimensional hidden information through the \textit{filtration} of a sequence of approximated topological spaces called simplicial complexes. Building a filtration can be seen as wearing a lens for examining the data set; different lenses, i.e. different filtrations, let extract different information from the topological space. In this paper we deal with two kinds of filtrations: the \textit{Lower star filtration} and the \textit{Vietoris - Rips filtration}

For the sake of completeness we introduce some formal concepts:
A \textit{topology} on a set $X$ is a subset $T \subseteq 2^X $ such that:
\begin{itemize}
\item[-] If $S_{1}, S_{2} \in T,$ then $S_{1} \cap S_{2} \in T$ (equivalent to: If $S_{1}, S_{2}, \dots, S_{n}  \in T$ then  $\cap_{i=1}^{n} S_{i} \in T$) 
\item[-] If $\{ S_{j} | j \in J  \} \subseteq T, $ then $\cup_{j \in J}S_{j} \in T.$
\item[-] $\emptyset,  X \in T.$ 
\end{itemize}

\begin{figure}[h!]
  \centering
  \includegraphics[width=0.45\textwidth]{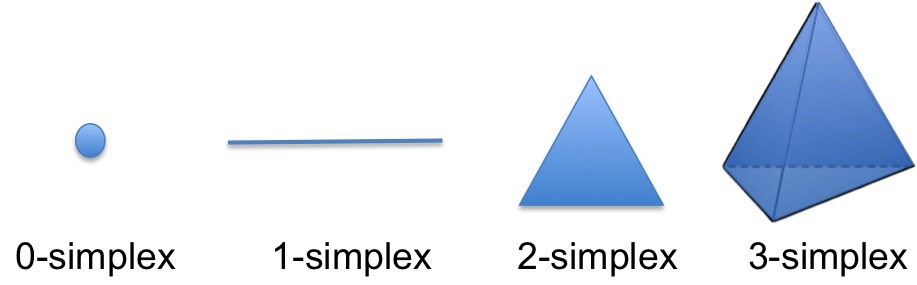}
  \caption[Topology by Rips]{Main simplices.}
  \label{fig:simplex}
\end{figure}

The pair $(X,T)$ of a set $X$ and a topology $T$ is a \textit{topological space}.
A {\it simplicial complex} is a kind of topological space constructed by the union of $n$-dimensional simple pieces in such a way that the common intersection of two pieces are lower-dimensional pieces of the same kind. More concretely,an abstract {\it simplicial complex} $K$ is composed by a set $K_0$ of $0-$simplices (also called vertices $V$,  that can be thought as points in $\mathbb{R}^n$)
and, for each $k\geq 1$, a set $K_k$ of $k-$simplices $\sigma=\{v_0, v_1, \dots, v_k\}$, where $v_i \in V$ for all $i\in \{0,\dots,k\}$, satisfying that:
\begin{itemize}
\item[-] each $k-$simplex has $k+1$ faces obtained removing one of its vertices;
\item[-] if a simplex $\sigma$ is in $K$, then all faces of $\sigma$ must be in $K$.
\end{itemize}
The underlying topological space of $K$ is the union of the geometric realization of its simplices: points for $0$-simplices, line segments for $1$-simplices, filled triangles for $2$-simplices,  filled tetrahedra for $3$-simplices and their $n$-dimensional counterparts for $n$-simplices (see Figure~\ref{fig:simplex}).
We only consider finite (abstract) simplicial complexes with finite dimension, i.e., there exists an integer $n$ (called the dimension of $K$) such that for $k>n$, $K_k=\emptyset$ and for $0\leq k\leq n$, $K_k$ is a finite set. 

A {\it filtration} on a simplicial complex $K$ is a collection of subcomplexes $\{K(t)|t\in\mathbb{R}\}$ of $K$ such that $K(t) \subset K(t')$ whenever $t\leq t'$. The filtration value of a simplex $\sigma \in K$ is the smallest $t$ such that $\sigma \in  K(t)$.
A {\it filtered simplicial complex} is a simplicial complex equipped with a filtration.

As the filtration is the lens for reading the data it is worth understanding what kind of information exactly we are looking for. The answer is: the topological invariants or if you prefer the $k$-dimensional holes in the data. During the filtration process $k$-dimensional holes appear and disappear and some of them persist for a certain filtration-time: the short-lived are considered noise while the long-lived are the invariants we are looking for, see \textsection{~\ref{sec:phom}}.
See \cite{hatcher2002algebraic} and \cite{Munkres} for an introduction to algebraic topology.

\subsection{How to build simplicial complexes from data}
\label{sec:fromdatatosimplicial}
\begin{figure}[h!]
  \centering
  \includegraphics[width=0.8\textwidth]{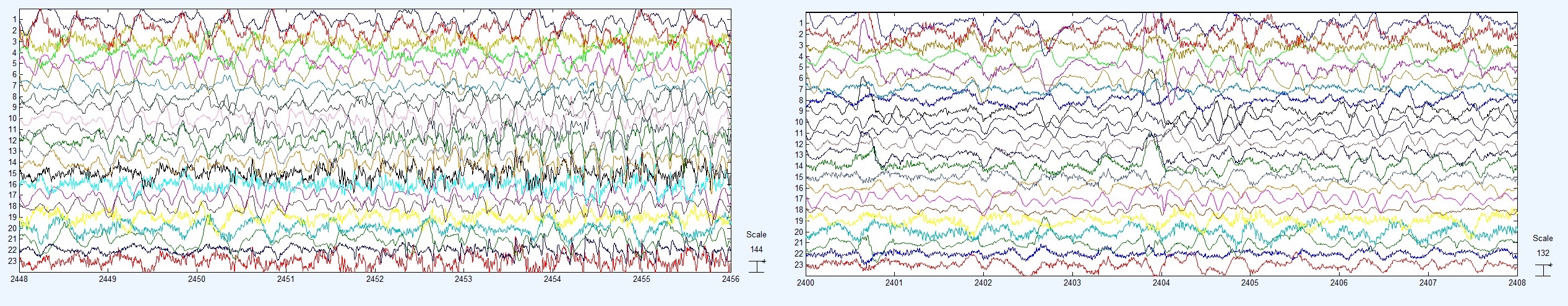}
  \caption[Epileptic time series]{Example of an electric signal recorded by a patient with epileptic seizures, on the left, and from a healthy patient, on the right.}
  \label{fig:epilessia}
\end{figure}

\begin{figure}[h!]
	\centering
		\includegraphics[scale=.24]{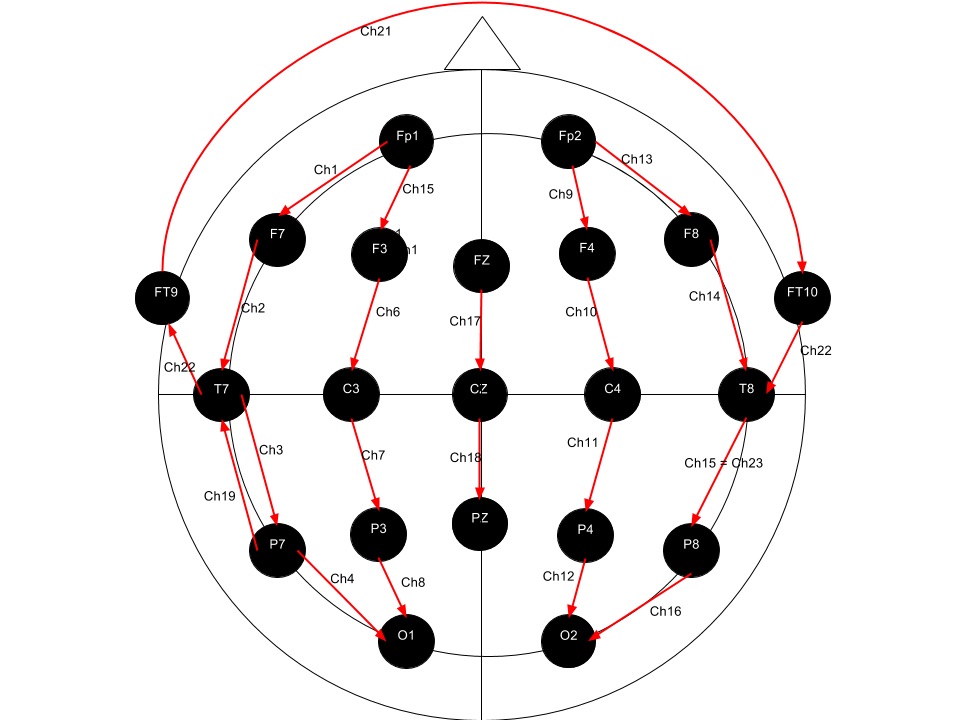}
	\caption{Graphical schema representing the positions of the sensors. The arrows correspond to the 23 potential differences that are recorded during the EEG.}
	\label{fig:positions}
\end{figure}

As we mention above, to choose how to build a filtration is very important because it reflexes how data points are converted into the simplicial complexes and thus converting set points into a topological space. Literature is plenty of techniques for obtaining simplicial complexes from data ~\cite{edelsbrunner2008persistent,carlsson2004persistence,binchi2014jholes} and each of them depends on the particular application we are interested in. In this paper data are represented by multivariate time series (see Figure\ref{fig:epilessia}). They are EEG signals of the \textit{PhysioNet} database and freely accessible via web\footnote{http://www.physionet.org}. For performing an EEG, electrodes are pasted at some key points on the patient’s head following some schemes. The 10-20 system is the internationally recognized method to apply the location of electrodes in EEG recording. The 10-20 refers to the fact that actual distances between electrodes are either $10\%$ or $20\%$ of front-back or right-left distance of the skull (see Figure~\ref{fig:positions}). The electrodes are assumed to record the electrical activity of the underlying cortical area that, in turn, is given by the interaction of the activity of all the neurons of that area. The EEG signals used in this study were collected at the Children's Hospital Boston, and they consists of EEG recordings from pediatric subjects with intractable seizures. Subjects were monitored for up to several days following withdrawal of anti-seizure medication in order to characterize their seizures and assess their candidacy for surgical intervention.
From the Physionet database we obtained 33 EEGs with at least one epileptic event and 33 records of EEG without epileptic events, by looking for patients with the same number of sensors and with recordings of the same length. Each record contains 23 one-dimensional signals with a sampling frequency of 256 Hz and of 921600 time points. 
Upon these data we build to kind of different complexes: the Piecewise complexes and the Vietoris-Rips complexes.

\subsection*{Piecewise complexes}
\label{subsec:pwc}
\begin{figure}[!ht]
\centering
\includegraphics[width=0.25\textwidth]{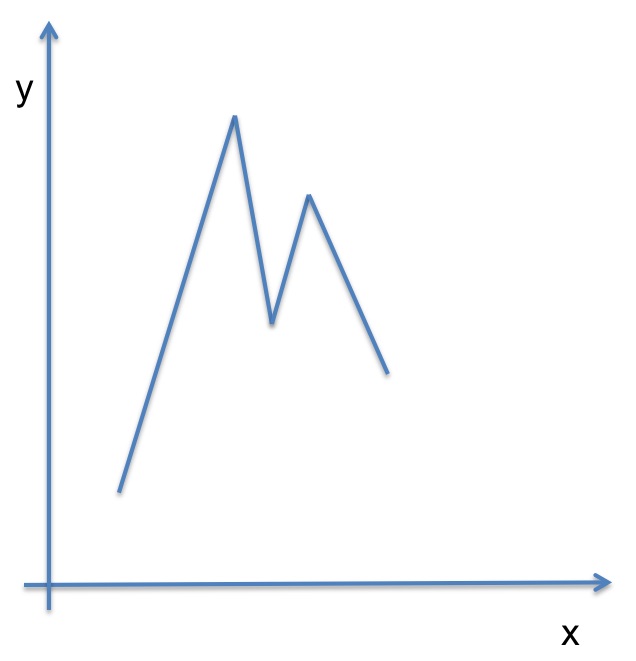}\\
\caption{Graphical representation of a piecewise linear function. The graph is composed of straight-line sections. }
\label{fig:PL}
\end{figure}

Piecewise linear function (PL) is a powerful mathematical tool largely used for approximating signals. The task of measuring the similarity among piecewise linear functions (PLs) is still an open issue and a solution is strongly required in machine learning methods. A piecewise linear function is a real-valued function defined on the real numbers (see Figure~\ref{fig:PL}).

In order to apply topological method to these functions, they must be equipped with a topology.  In~\cite{tdaLoccioni2}, the authors reported a new methodology that allows to associate a filtered simplicial complex to a PL. Then, Persistent Entropy, that is a topological summary is computed from the new simplicial complexes and it is used for the classification tasks of the topological spaces. Moreover, the authors have proven the stability theorem for the persistent entropy that is the formal method for comparing the persistent entropies of different simplicial complexes. Here we report on the theoretical methodology and one possible implementation. Given an unknown continuous signal $\tilde{f}:\mathbb{R}^n \to \mathbb{R}$, suppose that our input is the value of $\tilde{f}$ on  a finite set of points $S\subset \mathbb{R}^n$.
\begin{itemize}
\item[-] Let $K$ be a simplicial complex with real values specified at all vertices in $S$. E.g., if $S\subset \mathbb{R}$, then $K$ is a line subdivided in segments with endpoints in $D$.
\item[-] Using linear extension over the cells of $K$, we obtain a piecewise linear $(PL)$ function $f : K \to {\mathbb R}$ (being $f(u)=\tilde{f}(u)$ for  $u\in S$). It is convenient to assume that $f$ is {\it generic} by which we mean that the vertices have distinct function values.  To ensure unique values, $f$  may need to be perturbed.  One way of doing this is to add a linear ramp to $f$ (see \cite[page 1650]{vanessa}). We  can then order the vertices by increasing function value as $f(u_1) < f(u_2) < \dots < f(u_n)$.
\item[-] Now, the {\it lower star} of $u_i$ can be computed which  is the subset of simplices for which $u_i$ is the vertex with maximum function value, $$St\_ \,u_i = \{\sigma\in St\,u_i : x\in \sigma \Rightarrow f(x) \leq f(u_i)\}.$$ The considered filtration is the {\it lower star filtration} of $f$ (see \cite[page 135]{edelsbrunner2010computational}): $\emptyset=K_0\subset  K_1\subset \cdots \subset K_n = K$, in which $K_i$ is the union of the first $i$ lower stars. 
\end{itemize}
\begin{figure}[h!]
    \centering
         \includegraphics[width=0.6\textwidth]{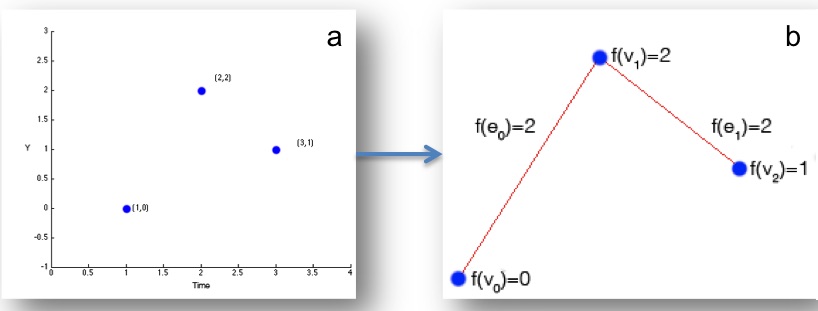}\\
     \caption{Graphical representation of the methodology that transforms a PL into a filtered simplicial complex.  From left to right: a) The input signal formed by three time points, respectively with coordinates: $(1,0),~(2,2),~(3,1)$. b) The filtered simplicial complex formed by three $0-$simplices: $\{v_0,v_1,v_2\}$ with filter values $f(v_0)=0, f(v_2)=1, f(v_1)=2$ and two $1-$simplices: $\{e_1, e_2\}$, with filter values $f(e_1)=f(e_2)=2$, so the filter-value set is $F=\{0,1,2\}$.}
\label{fig:construction}
\end{figure}
For practical purposes, the methodology explained above can be translated in the following algorithm
designed for analyzing a $2-$dimensional plot (see Figure~\ref{fig:construction}).
Suppose that the first coordinate of a point in $\mathbb{R}^2$ represents time. 
Given a signal $S\subset \mathbb{R}^2$:
\begin{itemize}
\item[-] order the points in $S$ respect to their first coordinate (i.e., order the points in $S$ by time);
\item[-] transform $S$ into a filtered simplicial complex: 
\begin{itemize}
\item[-]  each point of $S$ is a $0-$simplex with filter equal to  its second coordinate.
\item[-] Each pair formed by two consecutive points in $S$: $ (x_i,y_i)$ and $(x_{i+1},y_{i+1}) \in S$ where $x_i<x_{i+1}$, forms a $1-$simplex $\sigma$ with filter value $f(\sigma)=max\{y_i,y_{i+1}\}$. Note that the resulting filtration is obtained by presenting at the beginning the simplices formed with the lowest second coordinate (i.e., the  filter-value set $F$ is obtained by spanning the $Y-$axis in a upward direction).
\end{itemize}
The resulting filtration is a {\it lower start filtration}.
\end{itemize}

\subsection*{Vietoris-Rips complexes}
\label{subsec:vrc}
\begin{figure}[h!]
	\centering
		\includegraphics[scale=.3]{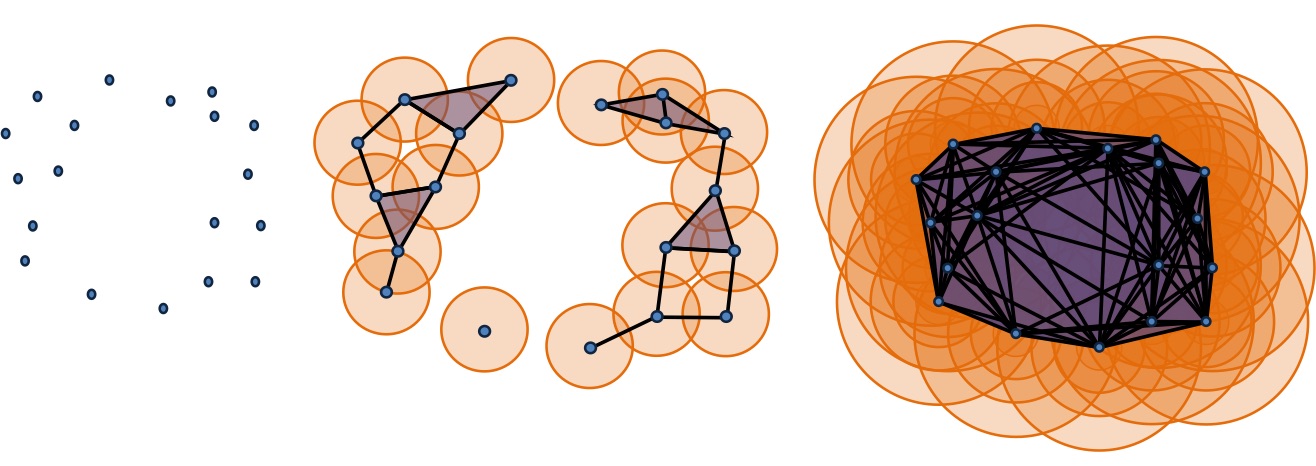}
	\caption{How to obtain Vietoris-Rips complexes from a PCD: starting from a PCD in a metric space, we surround each point with a sphere. All the spheres grow up simultaneously and equally. Each radius gives rise to new intersections. A new intersection of dimension k is equal to a k-1 simplices.}
	\label{figure_cech}
\end{figure}

Vietoris-Rips filtration is a versatile tool in topological data analysis and it is used for studying point cloud data (PCD).  More formally, Vietoris-Rips it is a sequence of simplices built on a metric space to add topological structure to an otherwise disconnected set of points. It is widely used because it encodes useful information about the topology of the underlying metric space. Two classical examples of abstract simplicial complexes are {\v C}ech complexes and  Vietoris-Rips complexes (see \cite[Chapter III]{edelsbrunner2010computational}).  Let $V$ be a finite set of points in $\mathbb{R}^n$.  The {\it {\v C}ech complex} of $V$ and $r$ denoted by {\it {\v C}}$_{r}(V)$ is  the abstract simplicial complex whose simplices are formed as follows. For each subset $S$ of points in $V$, form   a  closed ball of radius $r/2$ around each point in $S$, and include $S$ as a simplex of {\it {\v C}}$_{r}(V)$  if there is a common point contained in all of the balls in $S$. This structure satisfies the definition of abstract simplicial complex. The {\it Vietoris-Rips complex} denoted as $VR_{r}(V)$ is essentially the same as the {\v C}ech complex. Instead of checking if there is a common point contained in the  intersection of  the $(r/2)-$ball around $v$ for all $v$ in $S$, we may just check pairs
adding $S$ as a simplex of {\it {\v C}}$_{r}(V)$ if all the balls have pairwise intersections.  We have 
 $\mbox{\it {\v C}}_{r}(V) \subseteq VR_r(V)\subseteq \mbox{\it {\v C}}_{\sqrt{2} r}(V)$. See Fig.\ref{figure_cech}.
 
\subsection{How to analyze simplicial complexes: persistent homology}
\label{sec:phom}
A topological space is described by its own topological invariants. In case of the simplicial complex $K$ it might be studied by homology. {\it Homology} is an algebraic machinery that counts the number of ``n-dimensional'' holes in a topological space. Then, $K$ is described by the dimension of its homological groups. These invariants are called Betti numbers. The value $Betti_k$, where $k \in \mathbb{N}$, is equal to the rank of the $k$-th homology group of $K$. In particular, $Betti_0$ is the number of connected components, $Betti_1$ represents 2-dimensional holes, $Betti_2$ the voids in a 3-dimensional space, etc\dots. For instance, a $k$-dimensional empty sphere has all Betti numbers equal to zero except for $Betti_0$ = $Betti_k$ = 1. 
Persistent homology is the combinatorial counterpart of homology. It takes as input a filtered simplicial complex, it does not mind how the filtered simplicial complex has been obtained. Persistent homology describes how the homology of $K$ changes along filtration. A $k-$dimensional Betti interval, with endpoints $[t_{start}, t_{end})$, corresponds to a $k$-dimensional hole that appears at filtration time $t_{start}$ and remains until time $t_{end}$. We refer to the holes that are still present at $t= t_{max}$ as \textit{persistent topological features}, otherwise they are considered \textit{topological noise}~\cite{adams2011javaplex}. The set of intervals representing birth and death times of homology classes is called the {\it persistence barcode} associated to the corresponding filtration.  Instead of bars, we sometimes draw points in the plane such that a point $(x,y)\in \mathbb{R}^2$ (with $x< y$) corresponds to a bar $[x, y)$ in the barcode. This set of points is called  {\it persistence diagram}~\cite{zomorodian2005computing}. 
\begin{figure}[h!]
\centering
\includegraphics[width=0.8\textwidth]{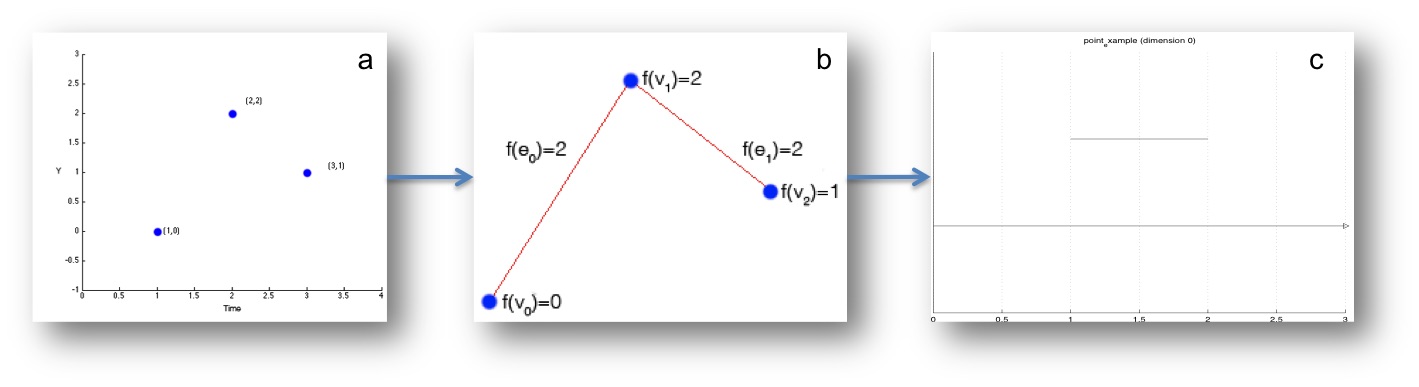}\\
\caption{Graphical representation of the methodology that transforms a PL into a filtered simplicial complex.  From left to right: a) The input signal. b) The corresponding filtered simplicial complex. c) The persistent barcodes. }
\label{fig:plF}
\end{figure}
In Figure~\ref{fig:plF} we represent the same simplicial complexes of Figure~\ref{fig:construction} but also equipped with its own persistent barcodes (box ``c''). The persistent barcodes: at $F=0$, there is only one topological feature corresponding to $v_0$; at $F=1$, $v_0$ is still in the space but also a new component is introduced and it corresponds to $v_2$; eventually for $F=2$, a new $0-$simplex is added to the topological space $(v_1)$ within the two $1-$simplices $e_1,~e_2$ where $e_1=\{v_0,v_2\}$ and $e_2=\{v_2,v_1\}$. From this filter value and successive, the space is described by only one persistent connected component, i.e. $\beta_0=1$. Visually there is only \textit{one infinite line in the barcode}. \\
Persistent homology does not limit to count the number of higher dimensional holes but it also returns the simplices that are involved in the holes. These simplices are called {\it generators}. Even if there is not a rigorous justification, there are a bunch of experimental observations advocating that these generators plays a crucial role for describing the data under analysis. For example, in \cite{merelli2015topological}, the authors used TDA for studying the human immune systems and they recognized that the generators of persistent holes are exactly the antibodies involved in the immune memory. In~\cite{lockwood2014topological}, the authors used the Vietoris-Rips complexes for analysing gene expression matrices and they discovered that the generators of the persistent holes are exactly the genes that are the responsible for certain pathologies.

\subsection{Statistics for persistent homology: persistent entropy}
\label{sec:pent}
In order to measure how much is ordered the construction of a filtered simplicial complex, a new entropy measure, the so-called \textit{the persistent entropy}, has been defined in~\cite{castiglione2014}. A precursor of this definition was given in~\cite{chintakunta2015entropy} to measure how different bars of the barcode are in length. Here  we recall the definition.

\vspace{0.5cm}
\noindent \textbf{Persistent entropy.}
Given a filtered simplicial complex $\{K(t) :  t\in F\}$, and the corresponding persistence barcode $B = \{a_i=[x_i , y_i) : i\in  I\}$, the \textit{persistent entropy} $H$ of the filtered simplicial complex is calculated as follows:
$$
H=-\sum_{i \in I} p_i  log(p_i)
$$
where $p_i=\frac{\ell_i}{L}$, $\ell_i=y_i - x_i$, and $L=\sum_{i\in I}\ell_i$.
Note that, when topological noise is present, for each dimension of the persistence barcode, there can be more than one interval, denoted by $[x_i~,~y_i)$, with $i \in I$. This is equivalent to say that, in the persistent diagram, the point $[x_i,y_i)$ could have multiplicity greater than $1$ (see \cite[page 152]{edelsbrunner2010computational}). 
In the case of an interval with no death time, $[x_i~,~\infty)$, the corresponding barcode $[x_i~,~m)$ will be considered, where $m = \max\{F\} + 1$.
Note that the maximum persistent entropy corresponds to the situation in which all the intervals in the barcode are of equal length. In that case, $H=\log n$ if $n$ is the number of elements of $I$. Conversely, the value of the persistent entropy decreases as more intervals of different length are present. The stability theorem for persistent entropy~\cite{tdaLoccioni2} lets to compare the entropies computed from the same simplicial complex but equipped with different filtrations.
 
For example, lets compute the persistent entropy for the piecewise simplicial complexes in Figure~\ref{fig:plF}.The persistent entropy of the space is computed as follows. The maximum filer value is $2$, so the symbol ``$\infty$'' representing the persistent topological feature is substituted with the value $m=3$. Then, the barcode is formed by two lines with lengths $\ell_1=1$ and $\ell_2=3$, respectively. So the total length $L=1+3=4$, for each line the probability is given by $p_1=1/4$ and $p_2=3/4$, and finally the persistent entropy is $H=0.5623$.

\section{A new topology based classifier of epilepsy}
\label{sec:classifier}
In this paragraph we describe a new methodology, that is based on Topological Data Analysis, for the analysis and classification of EEG signals. The methodology works on an EEG signal and it can be divided in three steps: in \emph{Step I} we preprocess the input. In \emph{Step II} we convert the input into a collection of piecewise complexes that are characterized by their persistent homology and persistent entropy. The input EEG signal is then characterized by the average of the persistent entropy of its components. The average value is used for the classification. In \emph{Step III} we perform two experiments. In the first experiment we aim to group brains by looking at their homological invariants, while in the second one we intend to identify the channels that characterize an EEG record. For this task we use the Vietoris-Rips complexes.\\
More formally, given an EEG signal $S$ that is composed by $n$ one-dimensional time varying signals , s.t. $S=\{s_1, s_2,\dots,s_n\}$\\

\emph{Step I - preprocessing}\\
For each $s_i \in S$:
\begin{enumerate}
\item[-] apply a lowpass filter for reducing the noise
\item[-] decrease the sampling rate but preserving its main features. This step is required for optimizing the complexity (both temporal and spatial) of the methodology\footnote{The conversion of a signal into a piecewise complex has a temporal complexity equal to $\Theta(n)$. Also, a piecewise simplicial complex has $n$ 0-simplices and $n-1$ 1-simplices, where $n$ is the number of temporal points within the input signal where $n$ is the number of temporal points within the input signal. The complexity of the computation of the persistent homology, in JavaPlex, has in the worst case a complexity of $O(m^3)$ and in the average case $\Theta(m^2)$ Where $m$ is the number of simplices. }. We dubbed the new decimated signal with $sd_i$.
\end{enumerate}

\emph{Step II - classification}\\
For each $sd_i$:
\begin{enumerate}
\item[-] obtain a piecewise complex $pw_i$, see Section~\ref{subsec:pwc}
\item[-] describe $pw_i$ with its persistent homology and its persistent entropy
\end{enumerate}
Now the input signal $S$ is represented by a vector of $n$ values of persistent entropy. Compute the average of this vector. The average value of persistent entropies is a one dimensional feature that is able to differentiate signal by looking at their shapes~\cite{tdaLoccioni2}.\\

\emph{Step III -  The role of generators}\\
In this task a brain is characterized by 23 points $p_i$ and each $p_i$ is within the $\mathbb{R}^n$ space equipped with a distance. For the sake of clarity, in this experiment $p_i = \{v_{i,t_1}, v_{i,t_2},\dots,v_{i,t_n}\}$, where $v_{i,t_n}$ is the voltage recorded by the $i-th$ sensor at time $t_n$. 
\begin{itemize}
\item[-] compute the pair-wise standardized euclidean distances among the  $23 \times n$ $sd_i$ signals
\item[-] from this metric space compute the Vietoris-Rips complex 
\item[-] compute the persistent homology of the Vietoris-Rips complex
\item[-] list the generators of higher dimensional holes (see Section~\ref{sec:phom}) and study their frequency. These generators are central for the dynamics recorded by the input signal $S$.
\end{itemize}

\section{Results and Final Remarks}
\label{sec:results}
In this section we report on the application of the methodology described in Section~\ref{sec:classifier} to the EEG signals. We implemented the methodology in Matlab and we used the JavaPlex package for computing the persistent homology~\cite{tausz2012javaplex}. \\

\subsection{Step I - preprocessing}
For the \emph{preprocessing step} we used the MATLAB command ``decimate''\footnote{http://it.mathworks.com/help/signal/ref/decimate.html}. The lowpass filter is characterized by the normalized cutoff frequency 0.8/r and passband ripple 0.05 dB. We tested the performance of our systems with a decimation factor of 10 and 100. The results are the same in both cases. In the following we report the results obtained with the analysis with decimation factor 100, then in our setting a EEG is composed by 23 channels $\times$ 9216 time points.

\subsection{Step II - classification}
\begin{figure}[h!]
	\centering
		\includegraphics[scale=.32]{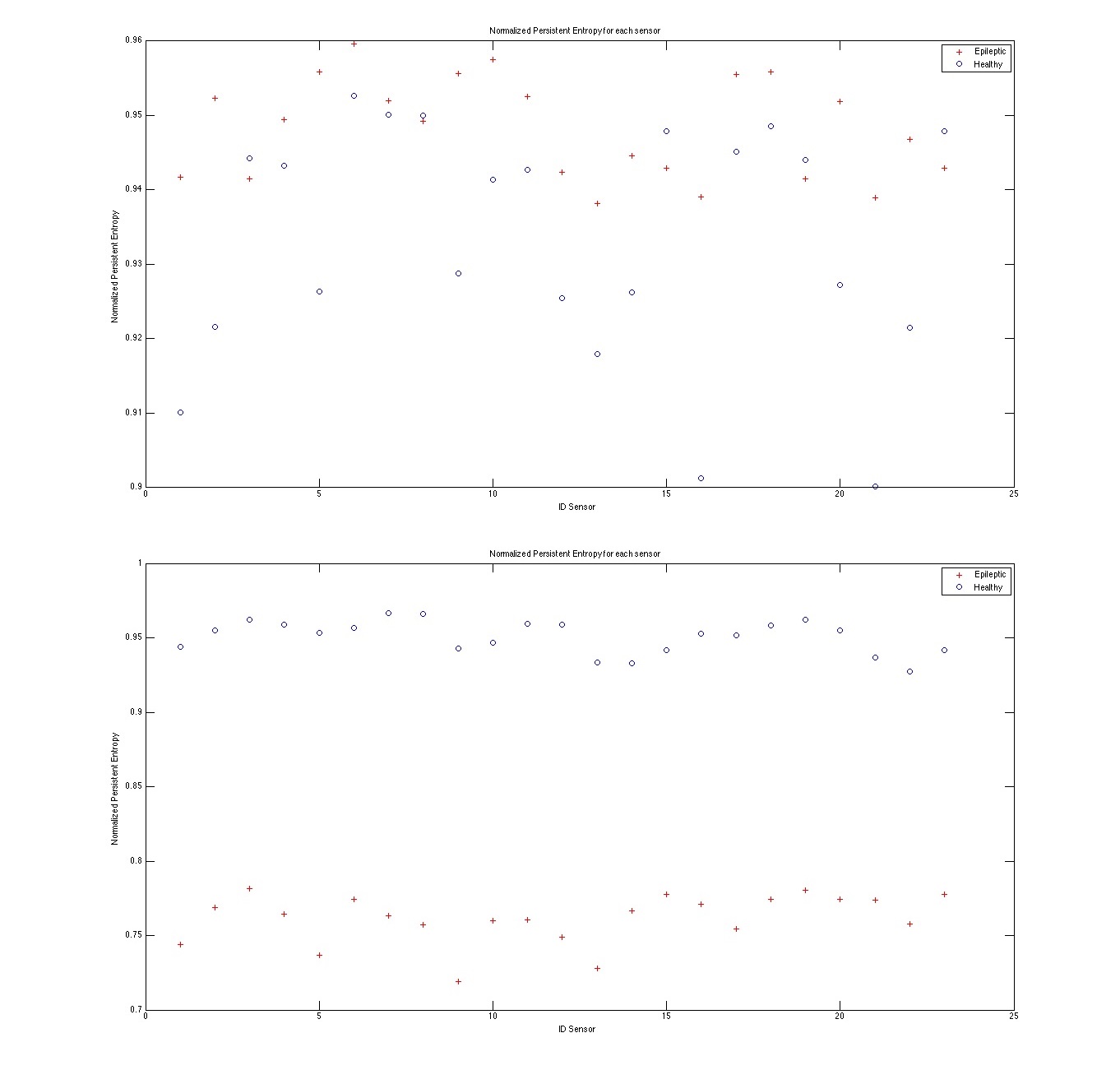}
	\caption{Normalized persistent entropy for the 23 signals within an EEG signal for  epileptic patient (red) and withing a EEG signal for healthy patient (blu). Top: weak separation between the healthy and the unhealthy patients. Bottom: strong separation between the healthy and the unhealthy patients.}
	\label{fig:entropies}
\end{figure}

We apply the methodology described in the paragraph~\ref{subsec:pwc} for converting the decimated signals into piecewise complexes, i.e. for creating the topological space, and the calculus of their persistent homology, the PE is derived. In Figure~\ref{fig:entropies} we plot the comparisons between two pairs of healthy an unhealthy patients. In the first pair (top) it is not possible to observe a strong separation among the values of the normalized persistent entropy, conversely the last pair (bottom) shows good separation.

\begin{figure}[t!]
	\centering
		\includegraphics[scale=.25]{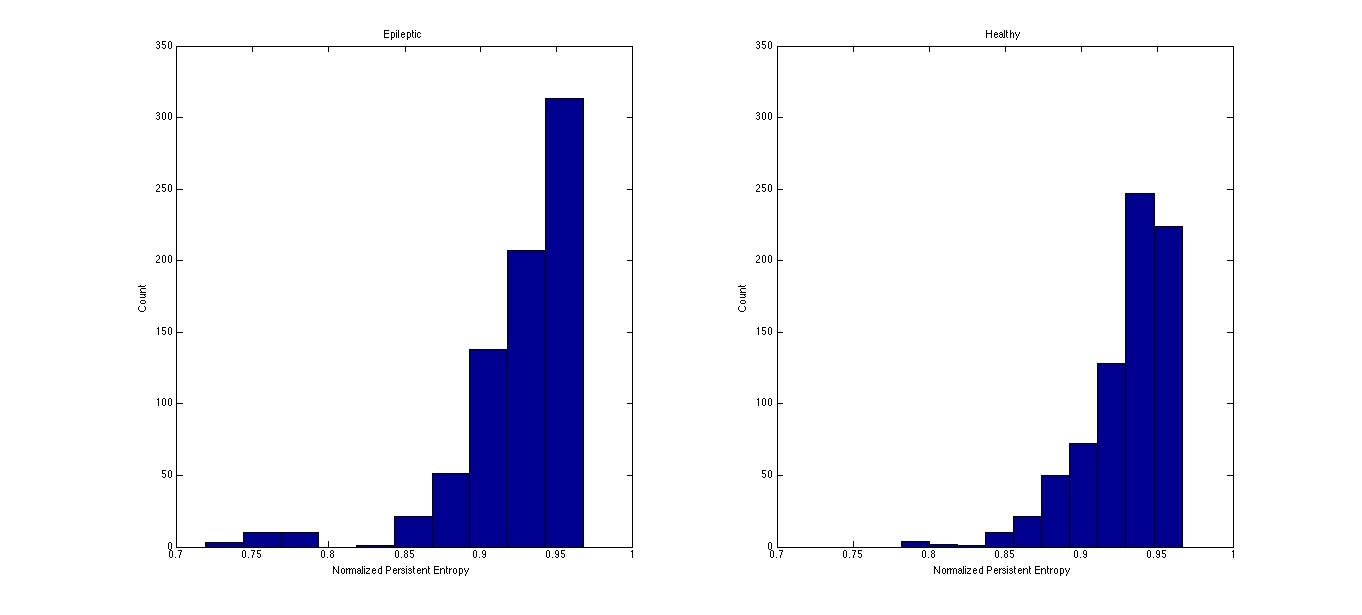}
	\caption{Histogram of the normalized persistent entropies for each sensor $(23\times 66)$ for the two classes of patients (left: epileptic, right: healthy). The elements are sorted into 10 equally spaced bins along the x-axis between the minimum and maximum values of normalized persistent entropy.}
	\label{fig:histogram}
\end{figure}

\begin{figure}[t!]
	\centering
		\includegraphics[scale=.25]{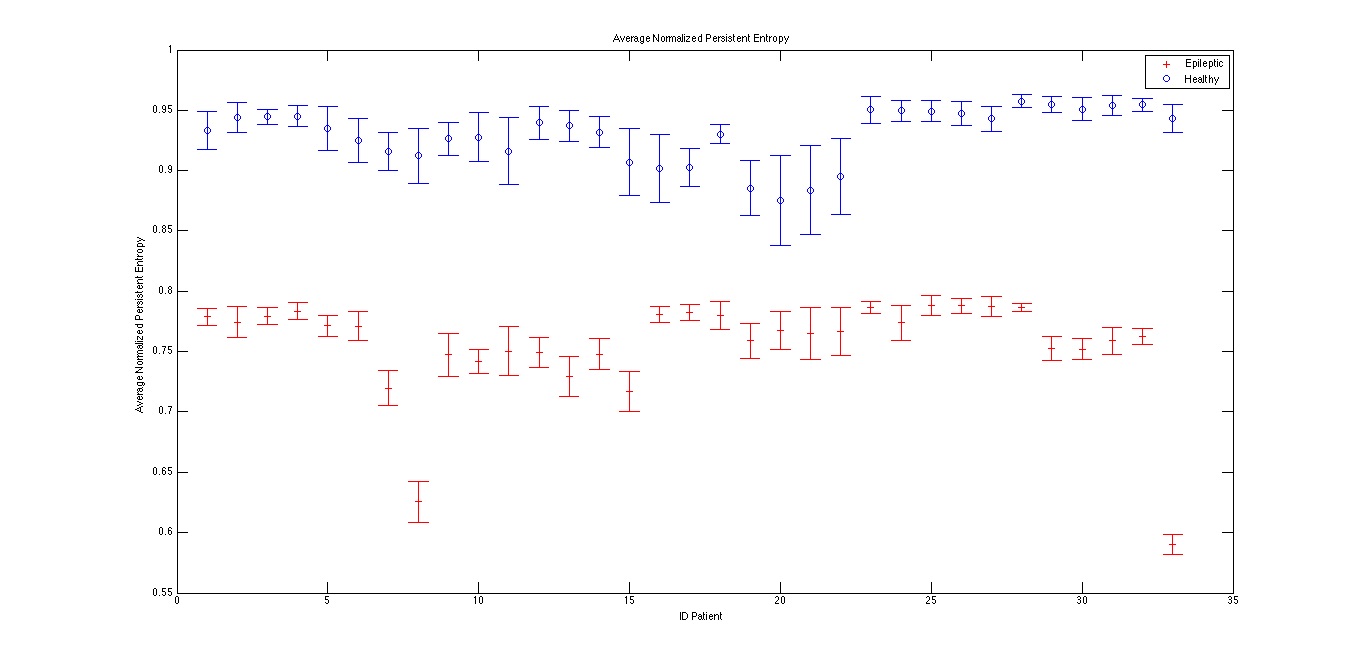}
	\caption{Average value of normalized persistent entropy over the 23 signals within the EEG for  epileptic patients (red) and  for healthy patients (blu). The error bars are the standard deviation.}
	\label{fig:average}
\end{figure}

\begin{figure}[t!]
	\centering
		\includegraphics[scale=.3]{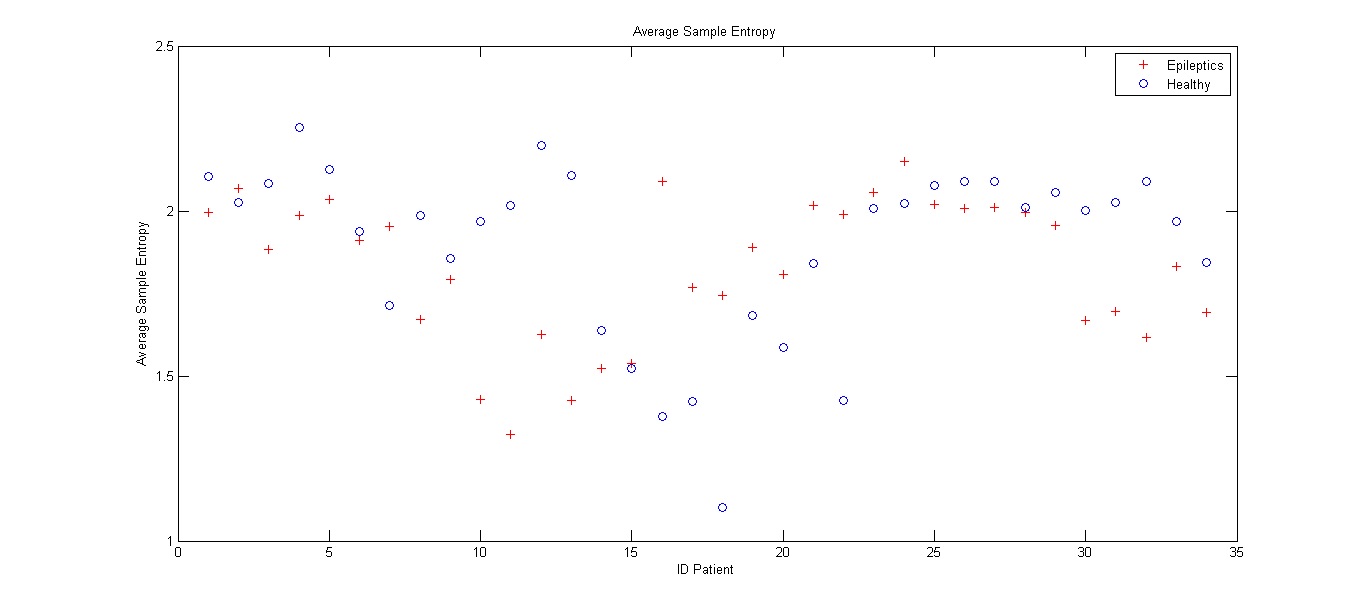}
	\caption{Average value of Sample entropy over the 23 signals within the EEG for  epileptic patients (red) and  for healthy patients (blu).}
	\label{fig:sampleent}
\end{figure}

From Figure~\ref{fig:histogram} we observe the class of epileptic patients is characterized by a peak of 313 elements in the range of normalized persistent entropies : $[0.942, 0.967]$ and center value 0.955, while the healthy patients are characterized by the biggest bin containing 247 elements with normalized persistent entropies values in $[0.930, 0.948]$ and center value 0.939. This first attempt does not immediately reveal any meaningful information for characterizing the two classes. Conversely, the global analysis of the EEG pinpoints out information that otherwise can not be detected. For this reason, instead of dealing with the analysis of the normalized persistent entropies for the single channel, we are focused on the study of the average value of the 23 values of persistent entropy. In Figure\ref{fig:average} we plot for each patient the average value of its normalized persistent entropy. It is well evident that there is a strong separation between the two populations.  We confirm the separation between the two classes by the statistical Wilcoxon test: p-value = 1.8346e-36, confidence interval = [1.6942, 1.9675].  We used the Wilcoxon test instead of the ``t-test'' because the two populations are not normally distributed. For having a comparison with a well establish method already used in the analysis of time series, we also calculated the Sample Entropy ~\cite{song2010new, richman2000physiological} of the data. The results can be seen in Figure \ref{fig:sampleent}. In this case sample entropy failed to separate the epileptic from the healthy. From the results of the PE we trained a supervised classifier. We randomly divided our dataset in two subsets: training and testing, of $70\%$ and $30\%$ respectively by applying a k-fold cross validation ($k=10$).

\begin{figure}[t!]
	\centering
		\includegraphics[scale=0.27]{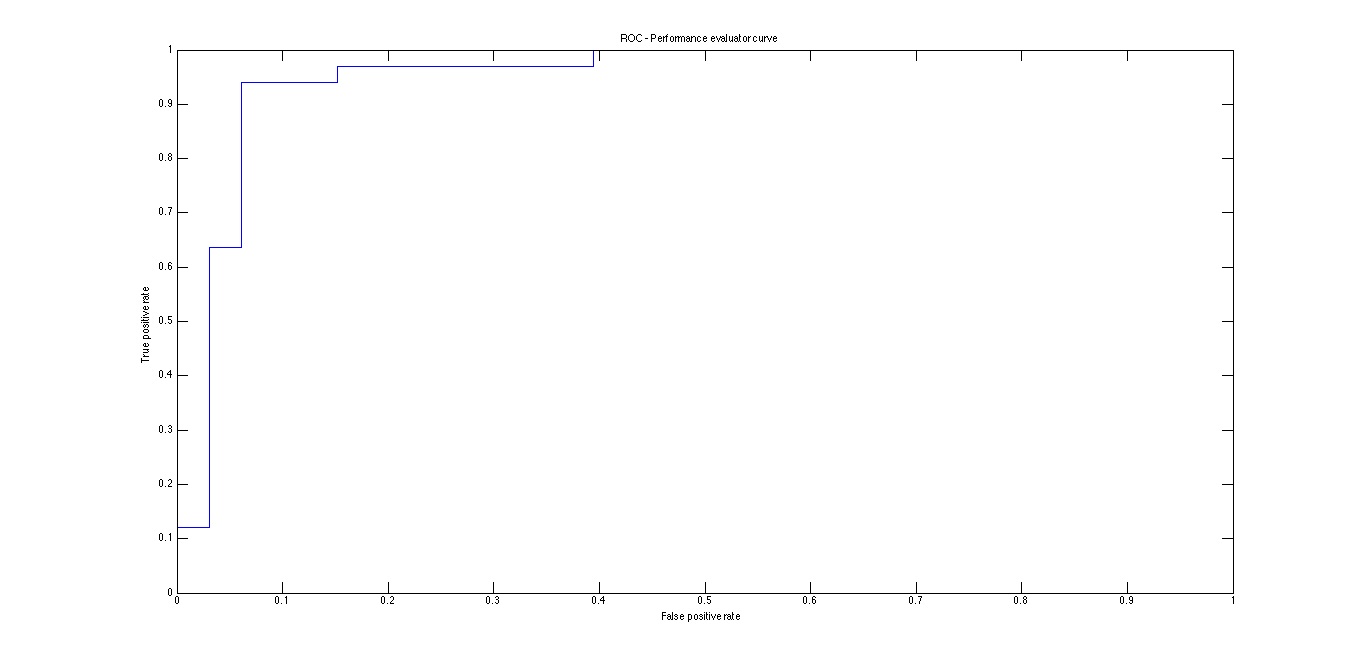}
	\caption{Receiver Operating Characteristic Curve with Area Under Curve of $97.2\%$.}
	\label{fig:roc}
\end{figure}

From the study of the ROC curve (see Figure~\ref{fig:roc}) we obtained that the best threshold for the separation of the two classes is: $\theta=0.8754$. It means, that given a new EEG if its normalized persistent entropy is less than $\theta$ then it is pathological, otherwise it is healthy. The area under curve (AUC) is $97.2\%$

\subsection{Step III -  The role of generators}
\begin{figure}[t!]
	\centering
		\includegraphics[scale=.5]{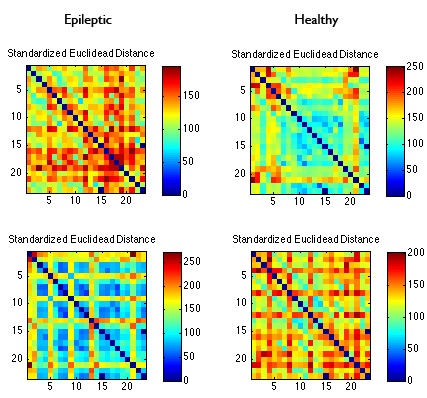}
	\caption{Four mosaics representing the pair-wise standardized euclidean distances among the 23 sensors of 4 EEG. Left: epileptic patients, right: healthy patients. From the analysis of the mosaics it is not possible to identify any patterns characterizing the two classes.}
	\label{fig:mosaics}
\end{figure}

\begin{figure}[t!]
	\centering
		\includegraphics[scale=.35]{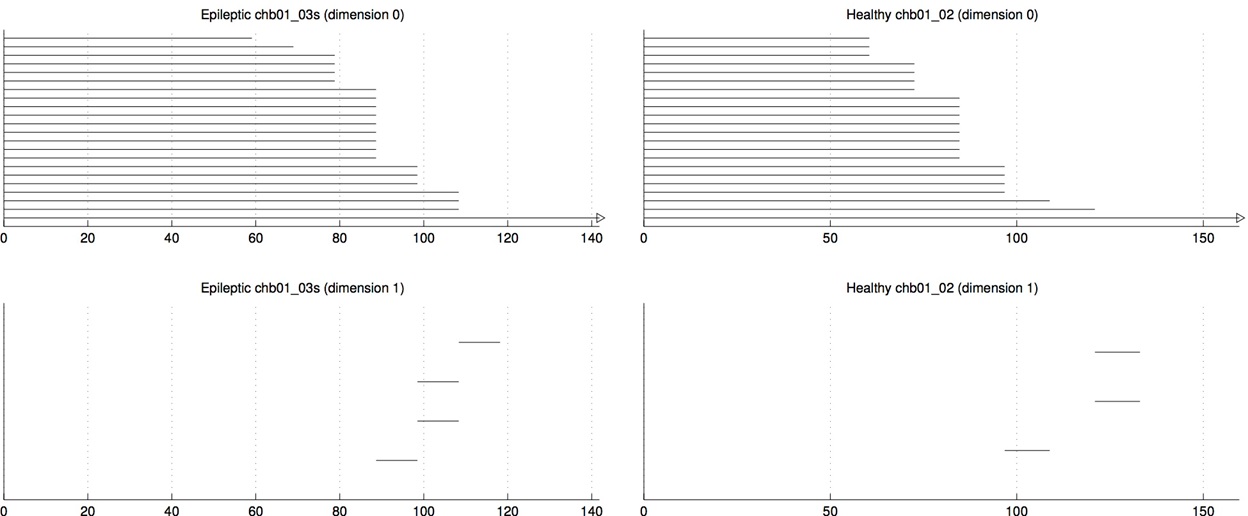}
		\caption{Example of persistent barcodes. Top: homological dimension 0 ($\beta_0$), bottom: homological dimension 1 ($\beta_1$). Left: epileptic patient, right: healty patient.}
	\label{fig:barcodesES}
\end{figure}

\begin{figure}[h!]
	\centering
		\includegraphics[scale=.25]{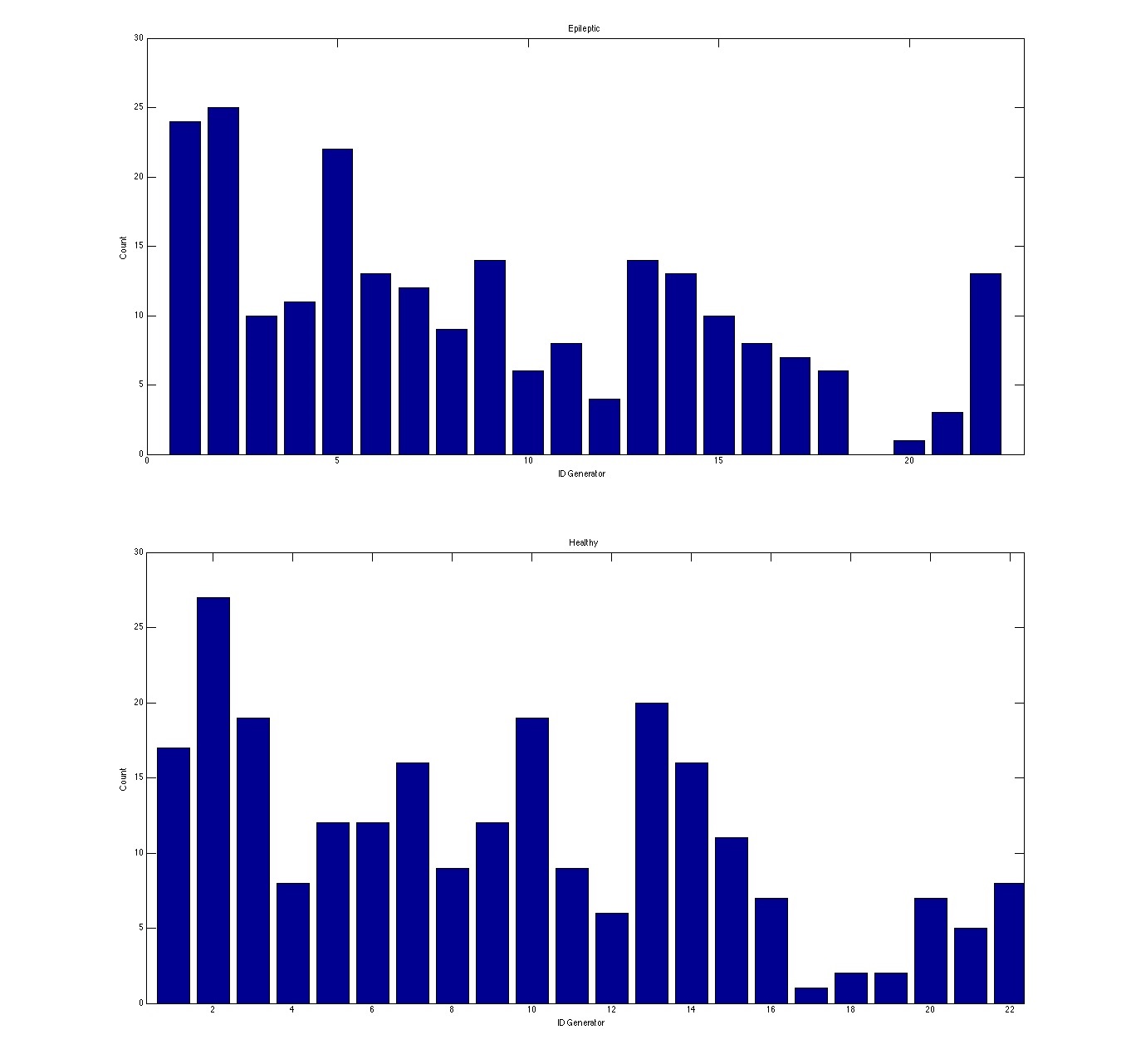}
		\caption{Bar plot of the generators belonging to higher dimensional holes for 33 epileptic patients (top) and 33 healthy patients (bottom).}
	\label{fig:histoGenerators}
\end{figure}

The last step of the methodology aims to identify the most important sensors for describing the brain activities recorded by an EEG analysis. We accomplish this task by analyzing the Vietoris-Rips associated to each patient~\cite{lockwood2014topological}. For each patient we compute the pair-wise standardized euclidean distance among the decimated signals (see Figure~\ref{fig:mosaics}). Standardized euclidean distance is able to give better performance when data set contains heterogeneous scale variables and it is defined in Equation~\ref{eq:sted}.
\begin{equation}
d(sd_1,sd_2)=\sqrt{\sum_{j=1}^J \Big(\frac{sd_{1,j}} {s_j} - \frac{sd_{2,j}} {s_j}\Big)^2 }
\label{eq:sted}
\end{equation}
Where $sd_{1,j}$ means the $j-th$ component of decimated signal 1, $s_j$ is the sample standard deviation of the $j-th$ variable.
From each metric space we compute the Vietoris-Rips and we analyzed their persistent homology and the generators of the persistent higher dimensional holes. In this setting the domain of the filter function coincides with the range of the pair-wise distances and we use as upper bound of the filtration the longest distances among sensors. In Figure~\ref{fig:barcodesES} we report two examples of persistent barcodes for one epileptic patient and for one healthy patient. We analyze all the 66 barcodes and for each patient we extract the number of $\beta_1$ (see Table~\ref{tab:bettiDistribution}) and the generators of the holes (see Figure~\ref{fig:histoGenerators}). 
From Table~\ref{tab:bettiDistribution} it is possible to deduce that there are less epileptic patients without $\beta_1$ than healthy patients (3 vs. 12).  It means that generally the epileptic patients are characterized by $\beta_1$. Unfortunately, this result can not be used for describing the two classes because the Wilcoxon test failed (p-value = 0.6946). However, the reader can recognize that the epileptic patients are principally characterized by 3 sensors, respectively with IDs: 1, 2 and 5 (see Figure~\ref{fig:positions} for matching the positions of the sensors). While the healthy patients are strongly characterized by sensor number 2 and also by sensors numbers: 1, 3, 7, 10 and 13. A potential interpretation is that, at least for this cohort, the epileptic events are on load of a few cortical columns, while conversely, healthy patients were stressed with stimuli that are recognized by more cortical columns. We argue this seminal result should be further investigated and if it will be validated it might be used for defining ad-hoc treatment.  

As future work we aim to exploit some authors’ results to characterize the synchronization of n-neurons interaction in term of oscillatory models (e.g. Kuramoto model) ~\cite{BartocciCMT09}. We are currently investigating the role of the simplices that generate the topological holes in the synchronization among the neurons. Our work intend to use the the extended Kuramoto model as backbone for the formal interpretation of our numerical results.

The final aim of the proposed approach in detecting the emerging behavior of epileptic seizures is to pave the ground for an automatic tool for supporting the clinicians during the diagnosis of epilepsy.

\begin{table}[!t]
\centering
\label{tab:bettiDistribution}
\begin{tabular}{||l|l|l||}
\hline

$\#\beta_i$ & Epileptic & Healthy      \\ \hline
0                & 3       & 12        \\ \hline
1                & 14      & 7         \\ \hline
2                & 5       & 8         \\ \hline
3                & 8       & 3         \\ \hline
4                & 1       & 1         \\ \hline
5                &         & 2         \\ \hline
\end{tabular}
\end{table}

\section*{Disclosure/Conflict-of-Interest Statement}
 
 The author thanks Dr. Giovanna Viticchi and Riccardo Ricciuti from AOU - Ospedali Riuniti di Ancona (Italy), for valuable discussions on medical aspects about epilepsy. The financial support of this paper was provided by the Future and Emerging Technologies (FET) program within the Seventh Framework Programme (FP7) for Research of the European Commission, under the FET-Proactive grant agreement TOPDRIM, number FP7-ICT-318121.  The authors declare that the research was conducted in the absence of any commercial or financial relationships that could be construed as a potential conflict of interest.

\end{document}